\documentstyle[preprint,aps]{revtex}

\newcommand{\be}{\begin{equation}}
\newcommand{\ee}{\end{equation}}
\newcommand{\ba}{\begin{eqnarray}}
\newcommand{\ea}{\end{eqnarray}}
\newcommand{\barr}{\begin{array}}
\newcommand{\earr}{\end{array}}

\begin{document}
\title{The Pairing interaction in nuclei: comparison between exact and approximate
treatments}
\author{J. Dukelsky$^{1}$, G. G. Dussel$^{2}$, J. G. Hirsch$^{3}$ and P. Schuck$^{4}$}

\address{
$^{\left( 1\right) }$Instituto de Estructura de la Materia,
C.S.I.C., Serrano 123, 28006 Madrid, Spain.\\
$^{\left( 2\right) }$Departamento de F\'{\i}sica Juan Jose
Giambiagi, Universidad de Buenos Aires. Pabell\'on 1, C. Universitaria, (1428) Buenos Aires, Argentina.\\
$^{\left( 3\right) }$ Instituto de Ciencias Nucleares, Universidad
Nacional Aut\'onoma de M\'exico, Apdo. Postal 70-543, M\'exico
04510 D. F.,
M\'exico.\\
$^{\left( 4\right) }$Institut de Physique Nucl\'eaire,
Universit\'e de Paris-Sud, F-91406 Orsay Cedex, France.}

\maketitle

\begin{abstract}
As a model for a deformed nucleus the many level pairing model
(picket fence model with ~100 levels) is considered in four
approximations and compared to the exact solution given by
Richardson long time ago. It is found that, as usual, the number
projected BCS method improves over standard BCS but that it is
much less accurate than the more sophisticated
many-body-approaches which are Coupled Cluster Theory ( CCT ) in
its SUB2 version or Self-Consistent Random Phase Approximation
(SCRPA).
\end{abstract}


\thispagestyle{empty}

\section{Introduction}

The importance of two nucleon pair correlations in the ground
state and low lying excited states of nuclei has been known for a
long time\cite{BS}. The application to nuclear systems\cite{BMP}
of the concepts used in the description of superconductivity in
solids was made immediately after the BCS theory has
appeared\cite{BCS}. During the sixties it was realized that the
pairing interaction was relevant in the description of two
particle transfer reactions in normal and superconducting
nuclei\cite{BB,BHR}. This interaction, which is usually thought to
represent the short range part of the bare nucleon interaction,
was treated by many authors in a phenomenological and schematic
way. Nevertheless, it has been found recently\cite{DZ} that the
pairing interaction is an important ingredient of the shell model
interaction derived from realistic forces and used in large scale
shell model calculations (the other two important ingredients
being the quadrupole-quadrupole and the monopole-monopole
interactions). It is also known\cite{TTD} that to preserve the
short range character of the force, it is necessary to use a large
number of shells.  Unfortunately the most simple theory for
pairing in finite nuclei, namely the mean field BCS approach, is
rather limited in its application, since particle number
fluctuations are very strong. Therefore more sophisticated
approaches such as particles number projection or the explicit
introduction of quantal fluctuations like BCS-QRPA approach
\cite{beha}  and other more elaborated theories have to be
considered.

In a series of papers between 1963 and 1968 Richardson \cite{Ri}
obtained the exact solution of the pairing hamiltonian providing
an analytic form for the eigenvalues and eigenvectors. These
papers have found a revival in the framework of ultrasmall
metallic grains \cite{delft}  where it was necessary to go beyond
the existing approximations to explain the disappearance of
superconductivity as the size of the grain \cite{DS1,DS2}
decreases. Subsequently, the exact solutions have been generalized
\cite{d1} and applied to other systems like Bose condensates
\cite{d2}, interacting boson models \cite{d3} and nuclear
superconductivity \cite{d4}.

The purpose of the present paper is to test on the exact solution
for a large scale case the precision of some well known
approximations like number projected BCS (PBCS) \cite{PBCS},
Coupled Cluster Theory (CCT) \cite{CCM},and Self-Consistent RPA
(SCRPA)\cite{SCRPA}. In reality the possibility of applying these
approximations depends on details of the nuclear residual
interaction. In general these approximations can deal in an
appropriated way with the long range part of the interaction, that
can be thought of, in a simplified way, as a particle-hole
interaction (as for example the quadrupole-quadrupole one), but
they may have problems in dealing with the short range part that
can be represented by the pairing interaction. The possibility or
convenience of using each of these methods depends on the strength
of the pairing interaction as well as the set of single particle
levels that is considered. For example for very strong pairing
(which is equivalent to a single shell) it is known that all the
particles participate in the ground state wave function, and
therefore one will need a quite large number of particle-holes
over the Hartree-Fock (HF) groundstate to describe properly the
paired state, on the other hand for a weak pairing interaction the
ground state wave function will be almost the HF one.

The paper is organized as follows.In Sect. II we present the
picket fence model and sketch the main steps for its exact
solution as given by Richardson. In Sect. III we outline the
various approximate methods to treat the model and in Sect. IV we
give the results together with a discussion. We end with the
conclusions in Sect V.
\bigskip

\section{\protect\bigskip The Model}
The picket fence model mimics superfluid correlations in a
deformed nucleus where the level density can be considered as more
or less constant and the levels are two-fold degenerate (for one
sort of nucleons). As mentioned in the introduction, the model has
been solved exactly in the early sixties by Richardson \cite {Ri}
for practically any number of levels. The latter feature makes the
model very interesting because one can treat situations, very
frequent in practice, which can not be mastered by ordinary
diagonalization techniques. For example we here will treat the
case of hundred particles distributed in hundred levels
corresponding to a dimension of the hamiltonian matrix of
$10^{29}$, well beyond any diagonalization technique. The model
has not been used very much in nuclear physics, probably because
of its schematic character. However, recently, its properties have
been exploited in rather great detail in the context of ultra
small superconducting metallic grains \cite {delft}. We here will
employ the model in order to assess the quality of commonly used
approximation schemes for nuclear pair correlations. We will consider
fermion \ creation $a_{\alpha m}^{\dagger }$ and annihilation
$a_{\alpha m}$ operators defined in a discrete basis labelled by
the quantum numbers $\left\{ \alpha m\right\} $. This basis can be
referred to the single particle states of an external potential,
 the single particle energies depend on the quantum numbers $\alpha
,m.$

The three operators

\begin{equation}
{n}_{\alpha m}=a_{\alpha m}^{\dagger }a_{\alpha m}\quad ,\quad
A_{\alpha m}^{\dagger }=a_{\alpha m}^{\dagger }a_{\alpha \overline{m}%
}^{\dagger }=\left( A_{\alpha m}\right) ^{\dagger }  \label{ope}
\end{equation}

close the commutator algebra

\begin{equation}
\left[ {n}_{\alpha m},A_{\beta n}^{\dagger }\right] =2\delta
_{\alpha \beta }\delta _{mn}A_{\alpha m}^{\dagger }~,~\left[
A_{\alpha m},A_{\beta n}^{\dagger }\right] =\delta _{\alpha \beta
}\delta _{mn}\left( 1-{n}_{\alpha m}\right)  \label{com1}
\end{equation}

In Eq. (\ref{ope}) the pair operator $A_{\alpha m}^{\dagger }$ creates a
pair of particles in the time reversal states $\left\{ \alpha m,\alpha \bar{m%
}\right\} $ where $a_{\alpha \overline{m}}^{\dagger }$ creates a particle in
the time reversed state of $a_{\alpha m}^{\dagger }$ . We will work with
nucleons interacting via a pure pairing force and for simplicity we will
represent by a single letter $k$ the quantum numbers $\alpha m$ (and when
there is no possibility of confusion it will represent the pair $\left\{
\alpha m,\alpha \bar{m}\right\} $). Therefore the Hamiltonian that we will
consider is
\begin{equation}
H=\sum_{k}\varepsilon _{k}\ n_{k}+G\sum_{k k'}A_{k}^{\dagger
}A_{k'} \label{H}
\end{equation}
where the $\varepsilon _{k}$ are the single particle energies.

The exact solution of this model has been obtained long ago by
Richardson \cite{Ri}. We will here briefly outline the method,
giving the equations to be used later on in the numerical
applications.

Richardson\cite{Ri} has shown that the exact eigenstates of the Hamiltonian (%
\ref{H}) with $M$ pairs can be written as

\begin{equation}
\left| \Psi \right\rangle =\prod_{i=1}^{M}B_{i}^{\dagger }\left| \varphi
_{\nu }\right\rangle  \label{psi}
\end{equation}
where there are $\nu $ unpaired nucleons. The state $\left|
\varphi _{\nu }\right\rangle $ describing the unpaired sector of
$\left| \Psi \right\rangle$ is defined by the action of the
operators $A$ and $n$ as

\begin{equation}
A_{k}\left| \varphi _{\nu }\right\rangle =0\quad ,\qquad n_{k}\left| \varphi
_{\nu }\right\rangle =\nu _{k}\left| \varphi _{\nu }\right\rangle  \label{fi}
\end{equation}
where $\nu _{k} = 1$ if there is one particle blocking the
state $k$ and $\nu_{k}=0$ elsewhere.

 The operator $B_{i}^{\dagger }$ in (\ref{psi}) creates a collective pair

\begin{equation}
B_{i}^{\dagger }=\sum_{k=1}^{\Omega }\frac{1}{2\varepsilon _{k}-E_{i}}%
A_{k}^{\dagger }  \label{B}
\end{equation}

where $\Omega $ is the number of single particle levels in the
valence space. The form of the amplitudes in (\ref{B}) were
suggested by the one pair diagonalization of the pairing
Hamiltonian (\ref{H}). The pair energies $E_{i}$ are unknown
parameters to be determined by the eigenvalue condition.

\begin{equation}
H\left| \Psi \right\rangle =E\left| \Psi \right\rangle  \label{eigen}
\end{equation}

After a long but straightforward derivation one arrives at the set
of $M$ nonlinear equations for the $M$ pair energies

\begin{equation}
1-2G\sum_{j\left( \neq i\right) =1}^{M}\frac{1}{E_{j}-E_{i}}%
+G\sum_{k=1}^{\Omega }\frac{\left( 1+2\nu _{k}\right)
}{2\varepsilon _{k}-E_{i}}=0  \label{richar}
\end{equation}
while the energy eigenvalue is

\begin{equation}
E=\sum_{i=1}^{M}E_{i}+ \sum_{k=1}^{\Omega}\varepsilon_{k} \nu_{k}
\label{ener}
\end{equation}

The pair energies $E_{i}$ are the roots of the set of $M$ coupled
equations (\ref{richar}). There are as many independent solutions
as states in the Hilbert space of $M$ pairs. The different
solutions, each one corresponding to an eigenstate of the pairing
hamiltonian, can be classified in the limit of $G \rightarrow 0$
as the different possible configurations of $M$ pairs in $\Omega$
levels, and then let them evolve adiabatically by solving the
equations (\ref{richar}) for increasing values of $G$ .

The occupation probabilities are obtained by means of the Gellman-Feynman
theorem, minimizing the energy with respect to the single particle energies

\begin{equation}
n_{k}=\frac{\partial E}{\partial \varepsilon _{k}}=\nu _{k}+\sum_{i=1}^{%
{\Omega}}\frac{\partial E_{i}}{\partial \varepsilon _{k}}
\label{occu}
\end{equation}

Differentiating (\ref{richar}) with respect to $\varepsilon _{k}$, the
occupation numbers can be expressed as

\begin{equation}
n_{k}=\nu _{k}+2\sum_{i=1}^{{\Omega}}\frac{\left( 1+2\nu _{k}\right) }{%
\left( 2\varepsilon _{k}-E_{i}\right) ^{2}}D_{i}  \label{occu1}
\end{equation}
where the $D_{i}$ should satisfy the system of equations

\begin{equation}
\left[ \sum_{k=1}^{\Omega }\frac{\left( 1+2\nu _{k}\right) }{\left(
2\varepsilon _{k}-E_{i}\right) ^{2}}+4\sum_{j\left( \neq i\right) =1}^{M}%
\frac{1}{\left( E_{j}-E_{i}\right) ^{2}}\right] D_{i}-4\sum_{j\left( \neq
i\right) =1}^{M}\frac{1}{\left( E_{j}-E_{i}\right) ^{2}}D_{j}=1
\label{occu2}
\end{equation}

\bigskip

The above equations are used to establish the exact solution with,
in the case considered here, a hundred levels with a hundred of particles.

\section{\protect\bigskip Approximate solutions}

We will study some approximations that are written in terms of
particular particle-hole excitations on a reference HF state. For
simplicity we will consider the case when the shells are half
filled, i.e. the number of pairs of particles $M$ will satisfy
$\Omega =2 M$. In the weak interaction limit the separation
between the energy levels is much greater than the gap. The
physics of this regime can be given in terms of the fluctuations
around the HF state

\begin{equation}
\left| HF\right\rangle =\prod_{h=1}^{M}A_{h}^{+}\left|
0\right\rangle \label{HF}
\end{equation}

where $h$ ($p$) refers to single particle states that are occupied
(unoccupied) in the limit $G=0$.

\subsection{\protect\bigskip Variational treatments}

We will first consider different variational treatments. The
simplest one is the standard BCS treatment. The next approximation
that we will consider is the number projected (before variation)
PBCS wave function where the ground state is assumed to be a
condensate of pairs of fermions. It is written as

\begin{equation}
\left| PBCS\right\rangle =\frac{1}{\sqrt{{Z}_{M, \omega}}}\left[
\Gamma ^{+}\right] ^{M}\left| 0\right\rangle \label{PBCS}
\end{equation}

where

\begin{equation}
\Gamma ^{+}=\sum_{k=1}^{\Omega }\lambda
_{k}A_{k}^{+}=\sum_{h=1}^{M} \lambda
_{h}A_{h}^{+}+\sum_{p=M+1}^{\Omega }\lambda _{p}A_{p}^{+}=\Gamma
_{h}^{+}+\Gamma _{p}^{+} \label{gamma}
\end{equation}

\begin{equation}
{Z}_{M, \Omega}=<0|\left[ \Gamma \right] ^{M} \left[ \Gamma
^{+}\right] ^{M} \left| 0\right\rangle \label{norm}
\end{equation}

and $\left| 0\right\rangle $ is the vacuum for the creation
operator of the nucleons. In general one minimizes the energy by
changing the variational parameters $\lambda _{k}$. In PBCS
$\lambda _{k}$ can be written in terms of the $v_{k}$ and $u_{k}$
parameters as $\lambda _{k}=\frac{v_{k}}{u_{k}}$ with
$v_{k}^{2}+u_{k}^{2}=1$ .

In Ref.\cite{DS2} the ground state energy was evaluated in terms of these $%
\lambda _{k}$ coefficients using the auxiliary quantities
\begin{equation}
Z_{N, \Omega}=<0|\left[ \Gamma \right] ^{N}\left[ \Gamma
^{+}\right] ^{N}\left| 0\right\rangle \label{s}
\end{equation}

\begin{equation}
S_{i}^{N}=<0|\left[ \Gamma \right] ^{N}A_{i}^{+}\left[ \Gamma
^{+}\right] ^{N-1}\left| 0\right\rangle \label{s1}
\end{equation}

\begin{equation}
Z_{ij}^{N}=<0|\left[ \Gamma \right] ^{N-1}A_{i}A_{j}^{+}\left[ \Gamma ^{+}%
\right] ^{N-1}\left| 0\right\rangle \label{z2}
\end{equation}

\begin{equation}
T_{ij}^{N}=<0|\left[ \Gamma \right] ^{N-2}A_{i}A_{j}\left[ \Gamma ^{+}%
\right] ^{N}\left| 0\right\rangle \label{t2}
\end{equation}

\bigskip and
\begin{equation}
\hat{S}_{i}^{N}=\frac{S_{i}^{N}}{Z_{N}};\hat{T}_{ij}^{N}=\frac{T_{ij}^{N}}{%
Z_{N}} \label{shat}
\end{equation}

The ground state energy is then written as

\begin{equation}
E_{gs}=2M\sum_{i}(2\epsilon _{i}-\mu )\lambda _{i}\hat{S}_{i}^{M}+G\sum_{ij}%
\lambda _{j}\hat{S}_{i}^{M}-GM(M-1)\sum_{ij}\lambda
_{i}^{2}\hat{T}_{ij}^{M} \label{egs}
\end{equation}

The auxiliary coefficients are determined by recurrence relations using the
fact that $Z_{0}=1$ ; $Z_{1}=\sum_{i}\lambda _{i}^{2}$ \ and $\hat{S}%
_{i}^{N}=\frac{\lambda _{i}}{Z_{1}}.$

The pair creation operator has two parts: one ($\Gamma _{p}^{+}$)
creates two particles above the Fermi sea while the other part
($\Gamma _{h}^{+}$ ) creates two particles below the Fermi sea. In
Ref.\cite{DS1} it is shown that if one defines the normalized
states

\begin{equation}
\left| K\right\rangle =\frac{1}{{Z}_{K, \Omega/2}}\left( \Gamma
_{p}^{+}\Gamma _{h}\right) ^{K}\left| HF\right\rangle \label{K}
\end{equation}

it is possible to write down the PBCS\ state as

\begin{equation}
|PBCS>=\sum_{K}\Psi _{K}^{PBCS}|K> \label{npbcs}
\end{equation}

where

\begin{equation}
\Psi _{K}^{PBCS}=\frac{\left( \left( \Omega /2\right) !\right) ^{2}}{\sqrt{%
{Z}_{\Omega /2, \Omega}{Z}_{\Omega /2,\Omega /2}}}\frac{{Z}%
_{K,\Omega /2}}{(K!)^{2}}={\cal A}_{\Omega }\frac{{Z}_{K,\Omega /2}}{%
(K!)^{2}}  \label{phibcs}
\end{equation}

and therefore the wave function can be written as

\begin{equation}
|PBCS>={\cal A}_{\Omega }\sum_{K} \frac{\left( \Gamma
_{p}^{+}\Gamma _{h}\right) ^{K}}{(K!)^{2}}\left| HF\right\rangle
\label{xxx}
\end{equation}
For details on this derivation see \cite{DS1}.

The variational parameters in this wave function are the amplitudes $\lambda
_{k}$. It must be taken into account that ${\cal A}_{\Omega }$ as well as
the operators $\Gamma _{p}^{+}$ and $\Gamma _{h}$ are well defined functions
of these parameters.

We also used another variational wave function with a structure
similar to the $exp(S_{2})$ type(see below), i.e.

\begin{equation}
|Exp>={\cal B}_{\Omega }\sum_{K} \frac{\left( \Gamma
_{p}^{+}\Gamma _{h}\right) ^{K}}{K!}\left| HF\right\rangle ={\cal
B}_{\Omega }\exp \left( \Gamma _{p}^{+}\Gamma _{h}\right) \left|
HF\right\rangle \label{var}
\end{equation}

In this case the dependence on the parameters $\lambda _{k}$
appears through the structure of $\Gamma _{p}^{+}$ and $\Gamma
_{h}$ and also in an indirect way in the normalization constant
${\cal B}_{\Omega }$.

\bigskip

\subsection{\protect\bigskip The Coupled Cluster Theory}

The CCT  has been proven in the past to be a highly performant
method for the calculation of correlation functions \cite{CCM}. It
has, however, never been tested for pairing model hamiltonians
which is an interesting study case because of its exact
solvability, even for very large number of particles.

The Hamiltonian of the picket fence model can be written in the
particle-hole basis as

\begin{equation}
H=\sum_{p}\varepsilon _{p}^{\prime }n_{p}+\sum_{h}\varepsilon
_{h}^{\prime }n_{h}-G\left\{ \sum_{p\neq p^{\prime
}}A_{p}^{\dagger }A_{p^{\prime }}+\sum_{h\neq h^{\prime
}}A_{h}^{\dagger }A_{h^{\prime }}+\sum_{ph}\left[ A_{p}^{\dagger
}A_{h}+A_{h}^{\dagger }A_{p}\right] \right\}  \label{ham}
\end{equation}

where

$\varepsilon _{p}^{\prime }=\varepsilon _{p}-G/2$ and $\varepsilon
_{h}=\varepsilon _{h}-G/2$

The unnormalized  CCT wave function in the SUB2 approximation
\cite{CCM} is

\begin{equation}
\left| \Psi \right\rangle =e^{S_{2}}\left| HF\right\rangle
~,~S_{2}=\sum_{ph}x_{ph}A_{p}^{\dagger }A_{h}  \label{CC}
\end{equation}
where the HF Slater determinant is given in (\ref{HF}). We have
stopped at the one p-pair one h-pair, i.e. at the SUB2 level for reasons given below. The aim of the CCT\ is to
determine the parameters $x_{ph}$ and the ground state energy.
Acting with the Hamiltonian on the wave function we have

\begin{equation}
H\left| \Psi \right\rangle =E\left| \Psi \right\rangle
=Ee^{S_{2}}\left| HF\right\rangle  \label{K1}
\end{equation}

The key point of the CCT\ is to multiply (\ref{K1}) with
$e^{-S_{2}}$ from the left. Then

\begin{equation}
e^{-S_{2}}H\left| \Psi \right\rangle =E\left| HF\right\rangle
\label{K2}
\end{equation}

\

Projecting on the HF bra

\begin{equation}
E=\left\langle HF\right| e^{-S_{2}}He^{S_{2}}\left|
HF\right\rangle , \label{K3}
\end{equation}

taking into account that

\begin{equation}
S_{2}^{\dagger }\left| HF\right\rangle =\left\langle HF\right|
S_{2}=0
\end{equation}

(\ref{K3}) is reduced to

\begin{equation}
E=\left\langle HF\right| He^{S_{2}}\left| HF\right\rangle
\label{ener0}
\end{equation}

Having in mind the form of the Hamiltonian (\ref{ham}), the
groundstate energy is

\begin{equation}
E=E_{HF}-G\sum_{ph}x_{ph}  \label{ener2}
\end{equation}

The amplitudes $x_{ph}$ are determined from the set of equations

\begin{equation}
\left\langle HF\right| A_{h}^{\dagger
}A_{p}e^{-S_{2}}He^{S_{2}}\left| HF\right\rangle =0  \label{ampli}
\end{equation}
which follows immediately after (\ref{K2})

Eq. (\ref{ampli}) can be expanded as

\begin{equation}
\left\langle HF\right| A_{h}^{\dagger }A_{p}\left( 1-S_{2}\right)
H\left( 1+S_{2}+S_{2}^{2}/2\right) \left| HF\right\rangle =0
\label{ampli1}
\end{equation}

The different terms are

\[
\left\langle HF\right| A_{h}^{\dagger }A_{p}H\left|
HF\right\rangle =-G
\]

\[
\left\langle HF\right| A_{h}^{\dagger }A_{p}\left( -S_{2}\right)
H\left| HF\right\rangle =-x_{ph}E_{HF}
\]

\[
\left\langle HF\right| A_{h_{1}}^{\dagger }A_{h_{2}}^{\dagger
}A_{h_{3}}A_{h_{4}}\left| HF\right\rangle =\left( 1-\delta
_{h_{1}h_{2}}\right) \left( \delta _{h_{1}h_{4}}\delta
_{h_{2}h_{3}}+\delta _{h_{1}h_{3}}\delta _{h_{2}h_{4}}\right)
\]

\[
\left\langle HF\right| A_{h}^{\dagger }A_{p}HS_{2}\left|
HF\right\rangle =2\left( \varepsilon _{p}^{\prime }-\varepsilon
_{h}^{\prime }\right) x_{ph}+2x_{ph}\sum_{h^{\prime }}\varepsilon
_{h^{\prime }}^{\prime }-G\left[ \sum_{p^{\prime }\left( \neq
p\right) }x_{p^{\prime }h}+\sum_{h^{\prime }\left( \neq h\right)
}x_{ph^{\prime }}\right]
\]

\[
\left\langle HF\right| A_{h}^{\dagger }A_{p}\left( -S_{2}\right)
HS_{2}\left| HF\right\rangle =Gx_{ph}\sum_{p^{\prime }h^{\prime
}}x_{p^{\prime }h^{\prime }}
\]

\[
\frac{1}{2}\left\langle HF\right| A_{h}^{\dagger
}A_{p}HS_{2}^{2}\left| HF\right\rangle =-G\left(
x_{ph}\sum_{p^{\prime }\left( \neq p\right) ,h^{\prime }\left(
\neq h\right) }x_{p^{\prime }h^{\prime }}+\sum_{p^{\prime }\left(
\neq p\right) ,h^{\prime }\left( \neq h\right) }x_{ph^{\prime
}}x_{p^{\prime }h}\right)
\]

And therefore it is possible to write the equation for $x_{ph}$ as
\[
2\left( \varepsilon _{p}^{\prime }-\varepsilon _{h}^{\prime
}\right) x_{ph}+2 Gx_{ph}\sum_{h^{\prime }}x_{ph^{\prime
}}+2Gx_{ph}\sum_{p^{\prime }}x_{p^{\prime }h}+2Gx_{ph}^{2}-G
\]
\begin{equation}
-G\sum_{p^{\prime }}x_{p^{\prime }h}-G\sum_{h^{\prime
}}x_{ph^{\prime }}+
 Gx_{ph}-G\sum_{p^{\prime
}h^{\prime }}x_{p^{\prime }h}x_{ph^{\prime }}=0 \label{CCM}
\end{equation}

This equation can be solved numerically.

\bigskip

\subsection{\protect\bigskip Self Consistent RPA}

The SCRPA for the Picket-Fence model has been developped in great
detail in Ref.\cite{hir,SCRPA0}. Here we give a brief summary.The
basic ingredients of the SCRPA approach in the particle-particle
channel are the two particle addition operator

\begin{equation}
A_{\mu }^{\dagger }=\sum_{p}X_{p}^{\mu }\
\overline{Q}_{p}^{\dagger }-\sum_{h}Y_{h}^{\mu }\ \overline{Q}_{h}
~,  \label{A}
\end{equation}
and the removal operator

\begin{equation}
R_{\lambda }^{\dagger }=-\sum_{p}Y_{p}^{\lambda }\overline{Q}%
_{p}+\sum_{h}X_{h}^{\lambda }\overline{Q}_{h}^{\dagger }~,
\label{R}
\end{equation}
where $\overline{Q}_{p}=A_{p}/\sqrt{1-\langle n_{p}\rangle }$ and $\overline{%
Q}_{h}=-A_{h}^{\dagger }/\sqrt{\langle n_{h}\rangle -1}$. Where
the expectation values are referred to the SCRPA vacuum defined as

\begin{equation}
A_{\mu }\left| SCRPA\right\rangle =R_{\lambda }\left|
SCRPA\right\rangle =0 \label{vac}
\end{equation}
and the collective RPA excitations are

\begin{equation}
\left| N+2\right\rangle _{\mu }=A_{\mu }^{\dagger }\left|
SCRPA\right\rangle \quad ,\quad \left| N-2\right\rangle _{\lambda
}=R_{\lambda }^{\dagger }\left| SCRPA\right\rangle \label{addrem}
\end{equation}

The equation of motion method applied to these operators leads
directly to the SCRPA equations
\begin{equation}
\left(
\begin{array}{cc}
A & B \\
-B & C
\end{array}
\right) \left(
\begin{array}{c}
X \\
Y
\end{array}
\right) =E\left(
\begin{array}{c}
X \\
Y
\end{array}
\right) ~,  \label{rpa}
\end{equation}
where
\begin{eqnarray}
A_{pp^{\prime }} &=&\langle
0|[\overline{Q}_{p},[H,\overline{Q}_{p^{\prime
}}^{\dagger }]]|0\rangle  \nonumber \\
&=&\delta _{pp^{\prime }}\left\{ 2 \varepsilon_p +G
+2{\frac{G}{1-\langle n_{p}\rangle }}\langle
(\sum_{p_{1}}A_{p_{1}}^{\dagger }+\sum_{h_{1}}A_{h_{1}}^{\dagger
})A_{p}\rangle \right\}
\nonumber \\
&&-G{\frac{\langle (1-n_{p})(1-n_{p^{\prime }})\rangle
}{\sqrt{(1-\langle
n_{p}\rangle )(1-\langle n_{p^{\prime }}\rangle )}}}~,  \nonumber \\
B_{ph} &=&\langle 0|[\overline{Q}_{p},[H,\overline{Q}_{h}^{\dagger
}]]|0\rangle =G{\frac{\langle (1-n_{p})(n_{h}-1)\rangle
}{\sqrt{(1-\langle
n_{p}\rangle )(\left\langle n_{h}\right\rangle -1)}}},  \label{scrpa} \\
C_{hh^{\prime }} &=&\langle
0|[\overline{Q}_{h},[H,\overline{Q}_{h^{\prime
}}^{\dagger }]]|0\rangle  \nonumber \\
&=&\delta _{hh^{\prime }}\left\{ -2 \varepsilon_h  +G
-2{\frac{G}{\langle n_{h}\rangle -1}}\langle A_{h}
(\sum_{p_{1}}A_{p_{1}}^{\dagger }+\sum_{h_{1}}A_{h_{1}}^{\dagger }
)\rangle \right\}
\nonumber \\
&&+G{\frac{\langle (n_{h}-1)(n_{h^{\prime }}-1)\rangle
}{\sqrt{(\langle n_{h}\rangle -1)(\langle n_{h^{\prime }}\rangle
-1)}}}~.  \nonumber
\end{eqnarray}
Since the amplitudes $X$ and $Y$ form a complete orthonormal set
of eigenvectors in (\ref{rpa}) one can invert the Bogoliubov
transformation of fermion pair operators (\ref{A},\ref{R}) and all
expectation values in (\ref{scrpa}) can be expressed in terms of
the RPA amplitudes and of the number operators expectation values
$\left\langle n_{p}\right\rangle $, $\left\langle n_{h}\right\rangle $%
, $\left\langle n_{p}n_{p^{\prime }}\right\rangle $, $\left\langle
n_{p}n_{h}\right\rangle $, and $\left\langle n_{h}n_{h^{\prime
}}\right\rangle $. For the particular case of the Picket Fence
models these expectation values can be calculated exactly within
the SCRPA approximation as shown in \cite{hir}. In this way the
SCRPA constitutes a closed set of equations without any further
approximation than the definition of the collective
operators $\left( \ref{A},\ref{R}\right) $ and the corresponding vacuum condition $\left( \ref{vac}%
\right) $.

Knowing these expectation values we can evaluate the SCRPA ground
state energy:

\begin{equation}
\langle H\rangle=\sum_{p}\varepsilon _{p}^{\prime }\langle n_{p}
\rangle +\sum_{h}\varepsilon _{h}^{\prime }\langle
n_{h}\rangle-G\left\{ \sum_{p\neq p^{\prime }}\langle
A_{p}^{\dagger }A_{p^{\prime }}\rangle+\sum_{h\neq h^{\prime
}}\langle A_{h}^{\dagger }A_{h^{\prime }}\rangle+\sum_{ph}\langle
A_{p}^{\dagger }A_{h}+A_{h}^{\dagger }A_{p}\rangle \right\}
\label{ham2}
\end{equation}

Assuming that the single particle energies $\varepsilon_{i}$ are all
equally spaced, separated by an energy gap $\varepsilon$, we have
$\varepsilon_{i}^{\prime } = \varepsilon_{i} -\varepsilon/2 +G/2 $.
In this case the SCRPA correlation energy is \be E^{SCRPA}_{corr} = \langle H
\rangle + \varepsilon M^{2} . \ee

\subsection{Results and discussion}

We will study the approximate descriptions of the pairing
interaction in the deformed nuclear region characterized by a
constant density of levels near the Fermi surface. This situation,
therefore,  can be represented by a set of equally spaced levels
with the appropriate density. We have used $100$ levels with a
constant level spacing of 300 keV and with 100 nucleons (half
filling). This represents typical values of the level density and
neutron numbers in the rare earth region ( $A\simeq 170$). As in
this region the gap has a value of the order of $\Delta \simeq
0.8$ MeV the physical value of the pairing interaction $G\simeq
0.1$ MeV. For this level density and number of particles the
critical pairing strength of the model in the BCS approximation
turns out to be $G_{c}\simeq 0.055$ MeV.

The aim here is to compare the quality of different approximations
to treat the pairing problem which are outlined in the text. We
display in Fig. 1 the ground state energy obtained using the
various methods discussed in the previous section (only the
correlation energy is displayed to isolate the effects due to the
interaction). All the correlations energies are given in terms of
the exact energy. Standard BCS approximation provides a rather
poor description. The numerical results do not appear in Fig. 1
because they are out of scale. A strong improvement over BCS is
obtained with the number projection before variation, i.e. the
PBCS procedure. Still quite a bit better works the Exp method for
moderate values of $G$ , described at the end of section III.A,
with the factorisable ansatz in the exponential. Both curves show
a typical structure: for small $ G $ there is a linear regime
which can be qualified as the perturbative regime. It is followed
by a part with negative curvature, characterized by precritical
fluctuations, before the superfluid regime develops after the
minimum. A detailed study of the two former regimes has been
performed in ref. \cite{prb}. The figure also shows a clear
indication that the PBCS approximation approaches the exact
groundstate energy in the large $G$ limit while this is not the
case for the Exp method. Both approximations underbind, as it
should be for a strictly variational theory in the sense of
Raleigh-Ritz. On the contrary CCT $(exp S_2)$ and SCRPA overbind
because neither CCT nor SCRPA in general correspond to a
Raleigh-Ritz theory. However, in absolute values both of the
latter theories work extremely well. It should be pointed out that
since SCRPA is a theory for two body correlation functions, we
only can go in CCT up to the $SUB2$ approximation, for
consistency. Going to higher approximations, we should also
include higher than two-body correlations and SCRPA and CCT would
not be on the same level of approximation.

We only have worked in the normal particle basis for CCT and SCRPA
and therefore the iterative solution of the eqs (\ref{CCM}) and
(\ref{scrpa}) did not converge any longer beyond $G/G_c \sim 1.3$.
We know from experience in other models \cite{SCRPA} that around
the mean field phase transition point one has to change to the
"deformed" basis which means to the quasiparticle basis in our
case. For PBCS and $Exp$ the error in the correlation energy in
the superfluid phase decreases for $G \geq G_c$ and therefore the
correlation energy has its maximal error in the transition region
as it is to be expected. For the picket fence model we have not
yet worked out the SCRPA in the superfluid phase and we are not
aware of any attempt to apply CCT in this regime. As mentioned
before, both curves in Fig. 1 stop at the point where we do not
find a numerical solutions of the corresponding equations any
more. We, however, conjecture that the end points of both curves
represent the maximal error and continuing the calculation in the
superfluid phase the error would start decreasing again. We see
that the errors in $exp S_2$ and SCRPA are, in the worst case,
only of $5 \%$ and $2 \%$ respectively. These errors are much
smaller than PBCS and $Exp$ which are of the order of $15 \% - 20
\%$. The very small errors of  $exp S_2$ and SCRPA is a very
satisfying result which confirms earlier positive results with
these theories for correlation functions in other cases. The
factor two improvement of SCRPA over $exp S_2$ for the correlation
energy in the transitional region has already been found in
another model study \cite{SCRPA} but this may be accidental.
Grossly speaking both methods are of similar characteristics and
accuracy for the correlation energy in the normal phase. The main
advantage we see in SCRPA is that excitation energies and
correlation functions are obtained simultaneously from the same
theory.  The SCRPA excitation energies also turn out to be very
accurate in the present model (see ref. \cite{hir}). In CCT the
excitation energies have to be constructed separately putting new
ingredients into the theory.

 In conclusion in this work
we have compared four methods for the calculation of energies in
the pairing case with parameters typical for deformed nuclei. This
study was performed in the picket fence model with a model space
of a hundred levels. The exact solution could be obtained owing to
the method proposed by Richardson long time ago, whereas a brute
force diagonalization is far beyond the limits of present
computers. We found that the $exp S_2$ and the SCRPA methods are
quite superior to the other variational methods in the normal
phase. The results obtained in this work might stimulate further
efforts to extend both approximations to the superfluid regime and
more realistic forces.

This work was supported in part by the the Spanish DGES under
grant \# BFM2000-1320-C02-02. GGD wants to thank the hospitality
of CSIS where this work started. This work has been supported in
part by the Carrera del Investigador Cient\'{i}fico y T\'{e}cnico,
by PID $N^{o}X-204/01$ of University of Buenos Aires, and by Conacyt,
M\'exico.

\vfill
\eject
\centerline{Figure Captions}

\begin{description}
\item[Figure 1:]  Ratio between the approximate and the exact correlation energies for
equally spaced levels as a function of the pairing strength for
the four approximations discussed in Section III.

\end{description}

\vfill
\eject

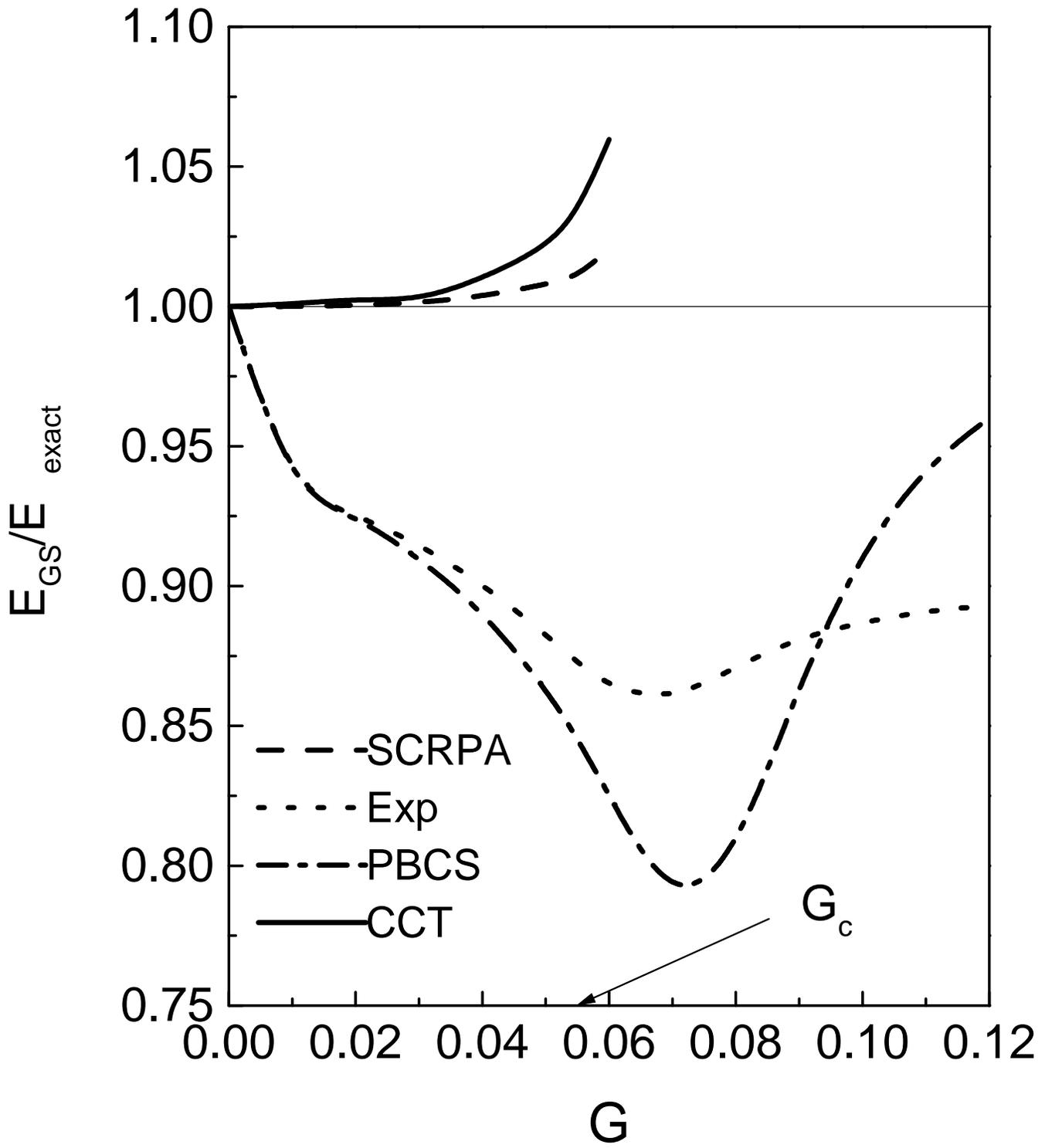
\end{document}